\newcommand{\cl}{\centerline}
\documentstyle[12pt]{article}
\begin{document}

\def\d{{\rm d}}
\begin{titlepage}
\hfill{CCUTH-94-06}\par
\setlength{\textwidth}{5.0in}
\setlength{\textheight}{7.5in}
\setlength{\parskip}{0.0in}
\setlength{\baselineskip}{18.2pt}
\vfill
\cl{\large{{\bf Applicability of Perturbative QCD to }}}\par
\cl{\large{{\bf $B\to D$ Decays}}}\par
\vskip 1.5cm
\cl{Hsiang-nan Li }
\vskip 0.5cm
\cl{Department of Physics, National Chung-Cheng University,}
\cl{Chia-Yi, Taiwan, R.O.C.}
\vskip 1.0cm
\cl{\today }
\vskip 4.0 cm
\cl{\bf Abstract}

We examine the applicability of perturbative QCD to $B$ meson decays
into $D$ mesons. We find that the perturbative QCD formalism,
which includes Sudakov effects at intermediate energy scales,
is applicable to the semi-leptonic decay $B\to D l\nu$, when
the $D$ meson recoils fast. Following this conclusion,
we analyze the two-body non-leptonic decays $B\to D\pi$ and $B\to DD_s$.
By comparing our predictions with experimental data,
we extract the matrix element $|V_{cb}|=0.044$.

\vfill
\end{titlepage}

\newpage
\cl{\large \bf 1. Introduction}
\vskip 0.5cm
Recently, perturbative QCD (PQCD) has been proposed to be an
alternative theory for the study of $B$ meson decays \cite{YS,LY,L}, which
complements the approach based on the heavy quark
effective theory (HQET) \cite{G} and the Bauer-Stech-Wirbel method
\cite{BSW}. The point is to include Sudakov effects \cite{BS},
which arise from the all-order summation of large radiative corrections
in the processes. It has been shown that these effects,
suppressing contributions due to soft gluon exchange, improve
the applicability of PQCD to exclusive processes around
the energy scale of few GeV \cite{LS}. The heavy $b$ quark
possesses a large mass scale located in the range of applicability
\cite{SHB}. The Sudakov factor for the heavy-to-light transition
$B\to \pi l \nu$ has been derived in \cite{LY}, and the perturbative
evaluation of the associated differential decay rate is found to be
reliable for the pion energy fraction above 0.3.

In this paper we shall investigate the applicability of the above PQCD
formalism to heavy-to-heavy transitions, concentrating on the
semi-leptonic decay $B\to D l\nu$.
Heavy quark symmetry \cite{IW} has been employed in the analysis of this
deacy \cite{N}, whose amplitude is written as
\begin{equation}
A(P_1,P_2)=\frac{G_F}{\sqrt{2}}V_{cb}{\bar \nu}\gamma_{\mu}(1-\gamma_5)l
\langle D (P_2)|{\bar c}\gamma^{\mu}b|B(P_1)\rangle\;,
\end{equation}
where $G_F=10^{-5}$ GeV$^{-2}$ is the Fermi coupling constant, and
$P_1$ ($P_2$) is the $B$ ($D$) meson momentum.
The transition matrix element
$M^\mu=\langle D|{\bar c}\gamma_\mu b|B\rangle$
can be expressed in terms of
a universal form factor $\xi$ in the heavy meson limit \cite{IW},
\begin{equation}
M^\mu=\sqrt{m_Bm_D}\xi(v_1\cdot v_2)(v_1+v_2)^\mu
\label{iw}
\end{equation}
with $m_B$ $(m_D)$ the $B$ $(D)$ meson mass. The velocities $v_1$
and $v_2$ are defined by the relations $P_1=m_Bv_1$ and $P_2=m_Dv_2$,
respectively. The form factor $\xi$, called the Isgur-Wise (IW) function,
depends only on the velocity transfer $v_1\cdot v_2$, and is normalized
to unity at zero recoil $v_1\cdot v_2=1$ in the limits
$m_B$, $m_D\to \infty$ \cite{IW}.

The IW function has been usually regarded as sensitive to
long-distance effects, and can not be calculated in perturbation theory.
For the behavior of $\xi$ above zero recoil,
there is only the model estimation
from the overlap integrals of heavy meson wave functions \cite{N3}.
In this work we shall show that PQCD can
give reliable predictions to $\xi$ in the region with large $v_1\cdot v_2$,
where the heavy quark symmetry can not provide any information of $\xi$.
We argue that the IW function is dominated by soft
contribution in the slow $D$ meson limit, at which the
heavy meson wave functions strongly overlap, and factorization theorems
do not hold. However, when the $D$ meson recoils fast, carrying the energy
much greater than $m_D$, the case is then similar to the $B\to\pi$ decays,
and PQCD is expected to be applicable \cite{KK}.

The above conclusion then indicates that two-body non-leptonic
decays such as $B\to D\pi$ and $B\to DD_s$ can be analyzed reliably in the
PQCD formalism.
The $B\to D$ decays have been studied \cite{KK,CM} based on the exclusive
PQCD theory developed by Lepage and Brodsky \cite{LB}. However,
these analyses are lack of quantitative justification for the perturbative
calculation, and are highly sensitive to the variation of the
heavy meson wave functions. Our predictions for the branching ratios
of these decay processess are comparable with those from
the standard PQCD in \cite{KK,CM}, and lead to the value 0.044 for
the Cabibbo-Kobayashi-Maskawa matrix element $|V_{cb}|$
by combining with experimental data \cite{KBS}. On one hand, we derive
the behavior of the IW function near the high end of $v_1\cdot v_2$.
On the other hand, the consistency of the extracted $|V_{cb}|$ with
its currently accepted value justifies the application of our PQCD
formalism to the semi-leptonic decays $B\to\pi l\nu$ \cite{LY}
and $B\to\pi\pi$ \cite{L}.

A model-independent extraction of the matrix element $|V_{cb}|$
has been obtained from the semi-leptonic decay $B\to D^* l\nu$ in the
framework of HQET \cite{N}. The value of $|V_{cb}|$ was read off
by extrapolating the experimental data
to the zero-recoil point, at which the IW function is known to be equal
to unity. In the present work, however, we must extract
$|V_{cb}|$ by studying the behavior of the IW function at the opposite
end of the velocity transfer, for which the PQCD analysis is reliable.
Hence, the two-body decays are good candidates.
Another possible method of extracting $|V_{cb}|$ has been proposed in
\cite{BDT}, in which a sum rule for the relevant structure function of
the inclusive semi-leptonic decay $b\to c$ was considered.

In section 2 we derive the factorization formulas for the
form factors involved in the $B\to D l\nu$ decay, including
the resummation of large radiative corrections to this
transition. Numerical analysis is shown
in section 3, along with the behavior of the IW function at large
velocity transfer. The
comparision of our predictions for the decays $B\to D\pi$ and
$B\to DD_s$ with experimental data is also made.
Section 4 contains the conclusions.
\vskip 2.0cm

\cl{\large\bf 2. Factorization}
\vskip 0.5cm

In this section we develop the factorization formula for the $B\to D l\nu$
decay. The lowest-order factorization for the transition matrix element
$M^{\mu}$ is shown in fig.~1, in which the bubbles represent
the $B$ and $D$ mesons, and the symbol $\times$ represents the
electroweak vertex where the lepton pair emerges.
The $b$ quark, denoted by a bold line, and its accompanying light quark
carry the momenta $P_1-k_1$ and $k_1$, respectively, which satisfy the
on-shell conditions $(P_1-k_1)^2\approx m_b^2$ and $k_1^2\approx 0$,
$m_b$ being the $b$ quark mass. We shall work in the rest frame
of the $B$ meson such that the nonvanishing components of $P_1$ are
$P_1^+=P_1^-=m_B/\sqrt{2}$. $k_1$ contains small amount of
transverse components ${\bf k}_{1T}$, and its minus component
defines the momentum fraction $x_1=k_1^-/P_1^-$.
The assignment of the momenta for the $D$ meson is similar, but with
$k_1$, $m_b$ and $x_1$ replaced by $k_2$, $m_c$ and
$x_2=k_2^+/P_2^+$, respectively, $m_c$ being the $c$ quark mass.

The expressions for the components of $P_2$ are more complicated.
At zero recoil the $D$ meson sits at rest with the $B$ meson, and we
have $P_2\propto P_1$. When the $D$ meson takes the maximum energy,
it moves fast and $P_2^+$ is much greater than
$P_2^-$. To show the relation between $P_2^+$ and $P_2^-$,
it is most convenient to express them in terms of the velocity transfer
$\eta=v_1\cdot v_2$. Solving the equations
$P_1\cdot P_2=\eta m_Bm_D$ and $P_2^2=m_D^2$, we obtain
\begin{eqnarray}
P_2^+&=&\frac{\eta+\sqrt{\eta^2-1}}{\sqrt{2}}m_D\;,
\nonumber \\
P_2^-&=&\frac{\eta-\sqrt{\eta^2-1}}{\sqrt{2}}m_D\;.
\label{ld}
\end{eqnarray}
The upper bound of $\eta$, corresponding to the maximum recoil
of the $D$ meson, is equal to $\eta_{\max}=(m_B/m_D+m_D/m_B)/2$.
It is easy to check from  eq.~(\ref{ld})
that $P_2^+=P_2^-=m_D/\sqrt{2}$ as
$\eta$ takes the minimum value 1, and $P_2^+/P_2^-=m_B^2/m_D^2\gg 1$ at
$\eta=\eta_{\max}$.

We then consider higher-order corrections to the basic factorization
picture. As analyzed before \cite{LY,BS,LS},
these corrections produce large
collinear logarithms, when the loop momentum is parallel to that
of a light quark, or large soft logarithms, when the loop momentum is much
smaller than the mass scale involved in the processes.
The two types of large corrections may combine to give double
logarithms. It has been shown that the double logarithms come from
two-particle reducible diagrams in physical (axial) gauge,
whose contributions are dominated by collinear enhancements for
fast light mesons,
and are dominated by soft enhancements for heavy mesons at rest \cite{LY}.
Therefore, they can be absorbed into the corresponding wave functions,
which involve similar dynamics.
The all-order summation of the double logarithms in light meson wave
functions, such as a pion, has been performed in \cite{LS}. The
resummation technique \cite{CS} has been extended to the case of heavy mesons
in \cite{LY}. Combining the above
results, we have derived the Sudakov factor for the heavy-to-light
transition $B\to\pi l\nu$ \cite{LY}.

The analysis of Sudakov corrections to fig.~1 is more complicated
compared to that of the decay $B\to\pi l\nu$. Due to the dominance
of soft contribution near the low end of $\eta$,
we concentrate only on the large $\eta$ region. In this region
radiative corrections on the $D$ meson side involve three
scales, $P_2^+\gg m_D\gg k_{2T}$. Note that all the previous studies of
resummation involve only two scales, for example, $P^+$ and $k_T$ in the
pion case, and $m_B$ and $k_T$ in the $B$ meson case.
The three scales produce various large logarithms of
$P_2^+/k_{2T}$, $P_2^+/m_D$ and $m_D/k_{2T}$, which complicate
their organization. As a noive approximation, we keep only the
largest one proportional to $\ln(P_2^+/k_{2T})$. The neglect of those
logarithms containing $m_D$ is equivalent to the neglect of $P_2^-
\ll P_2^+$ in the analysis of radiative corrections to the $D$ meson
wave function. The $D$ meson is then regarded as light in the large
$\eta$ region, and the corresponding Sudakov factor for the decay
$B\to Dl\nu$ can be approximated by that for the heavy-to-light transitions.

The factorization formula for $M^\mu$ in the transverse configuration space,
with radiative corrections taken into account, is written as
\begin{eqnarray}
M^\mu&=&\int_0^1 \d x_{1}\d x_{2}\int
\frac{\d^2 {\bf b}_1}{(2\pi)^{2}}\frac{\d^2 {\bf b}_2}{(2\pi)^{2}}
\,{\cal P}_D(x_{2},{\bf b}_2,P_{2},\mu)
\nonumber \\
& &\times \,{\tilde H}^\mu(x_1,x_2,{\bf b}_1,{\bf b}_2,m_B,m_D,\mu)
\,{\cal P}_B(x_{1},{\bf b}_1,P_{1},\mu)\; ,
\label{fbd}
\end{eqnarray}
in which both the $B$ and $D$ meson wave functions,
${\cal P}_B$ and ${\cal P}_D$, contain the evolution
from the resummation of double logarithms performed
in axial gauge. We have introduced the conjugate variable
$b_1$ $(b_2)$ to denote the separation between the two valence
quarks of the $B$ $(D)$ meson.
We shall employ the approximation $m_b\approx m_B=5.28$ GeV and
$m_c\approx m_D=1.87$ GeV in eq.~(\ref{fbd}). ${\tilde H}^{\mu}$
is the Fourier transform of the hard scattering amplitude $H^{\mu}$
to $b$ space. $\mu$ is the renormalization and factorization scale.

Note that in the evaluation of $H^\mu$ we neglect those terms proportional
to $k_1^+$ and $k_2^-$ in the hard scattering amplitude following
the kinematic ordering $k_1^+\sim k_2^-\ll k_1^-\sim k_2^+$, which is valid
in the large $\eta$ region. For example, the gluon propagator in the
lowest-order diagram is written as
\begin{equation}
\frac{1}{(k_1- k_2)^2+i\epsilon}\approx
\frac{-1}{2k_1^-k_2^++({\bf k}_{1T}-{\bf k}_{2T})^2}\;,
\label{ng}
\end{equation}
where ${\bf k}_T$ serves as the infrared cutoff of the Sudakov corrections.
Once the approximation is made, the $k_1^+$ and $k_2^-$ dependences, appearing
only in the $B$ and $D$ meson wave functions, respectively,
are integrated to give eq.~(\ref{fbd}).

As stated above, near the high end of $\eta$ the Sudakov factor $\exp(-S)$
for the decay $B\to Dl\nu$, which groups the large logarithmic corrections
in ${\cal P}_B$, ${\cal P}_D$ and ${\tilde H}^\mu$, can be approximated by
that for the heavy-to-light transition
derived in \cite{LY}, with the exponent $S$ given by
\begin{eqnarray}
S(x_i,b_i,m_B,m_D)&=&s(x_1,b_1,P_1^-)+
\sum_{x=x_2,\;\;1-x_2}s(x,b_2,P_2^+)
\nonumber \\
& &-\frac{1}{\beta_{1}}\left[\ln\frac{\ln(t/\Lambda)}{-\ln(b_1\Lambda)}+
\ln\frac{\ln(t/\Lambda)}{-\ln(b_2\Lambda)}\right]\;,
\label{sdb}
\end{eqnarray}
where $t$ is the largest mass scale associated with the hard gluon,
and will be specified later. The first term in eq.~(\ref{sdb}) comes from
the resummation of redicible corrections to the heavy meson wave function
\cite{LY}.
The value of $\Lambda\equiv \Lambda_{\rm QCD}$ will be set to 100 MeV below.
The complete expression for the factor $s(x,b,Q)$, including the leading
and next-to-leading logarithms, is exhibited in Appendix.
It is observed that $\exp(-S)$ decreases quickly in the large $b_i$
region and vanishes as $b_i>1/\Lambda$. Therefore, the long-distance
contributions are suppressed, and the perturbative
calculation becomes relatively reliable.

One may wonder whether the resummation of large radiative corrections
can improve the applicability of PQCD near the low end of $\eta$.
If we recognize that the $D$ meson is regarded as a heavy meson in this
region, and is dominated by similar dynamics to that of the $B$ meson,
the Sudakov factor for the decay $B\to Dl\nu$ can be taken as
the combination of the expressions for heavy mesons \cite{LY} at two
different mass scales, $m_B$ and $m_D$. The Sudakov exponent $S$ is
then written as
\begin{eqnarray}
S(x_i,b_i,m_B,m_D)&=&s(x_1,b_1,P_1^-)+
s(x_2,b_2,P_2^+)
\nonumber \\
& &-\frac{1}{\beta_{1}}\left[\ln\frac{\ln(t/\Lambda)}{-\ln(b_1\Lambda)}+
\ln\frac{\ln(t/\Lambda)}{-\ln(b_2\Lambda)}\right]\;.
\label{sda}
\end{eqnarray}
Obviously, it is not expected
that our perturbative analysis with the above Sudakov suppression
becomes self-consistent. The virtuality of the
hard gluon in fig.~1 diminishes as $x_1$ and $x_2$ are both small,
which leads to a large running coupling constant $\alpha_s$. However,
this non-perturbative region is not strongly suppressed by the Sudakov
factor in eq.~(\ref{sda}). It is the extra exponent $s(1-x_2,b_2,P_2^+)$
in eq.~(\ref{sdb}) that can provide necessary suppression in the small
$x_2$, or large $1-x_2$, region.

Having factorized all the large logarithms into the Sudakov factor,
we can then compute the hard scattering amplitude $H^\mu$ of the
$B\to D l\nu$ decay to lowest order of $\alpha_s$.
{}From fig.~1a we have
\begin{eqnarray}
H^{(a)\mu}&=&
{\rm tr}\left[\gamma_\alpha\frac{\gamma_5(\not P_2+m_D)}{\sqrt{2N_c}}
\gamma^\mu\frac{\not P_1-\not k_2+m_B}{(P_1-k_2)^2-m_B^2}
\gamma^\alpha\frac{(\not P_1+m_B)\gamma_5}{\sqrt{2N_c}}\right]
\nonumber \\
& &\times \frac{-g^2N_c{\cal C}_F}{(k_1-k_2)^2}
\nonumber \\
&=&\frac{16\pi \alpha_s{\cal C}_F[m_Bm_D-x_2\zeta_1m_D^2]}
{[x_1x_2\zeta m_Bm_D+({\bf k}_{1T}-{\bf k}_{2T})^2]
[x_2\zeta m_Bm_D+{\bf k}_{2T}^2]}P_1^\mu
\nonumber \\
& &+\frac{16\pi \alpha_s{\cal C}_F[m_B^2+x_2\zeta_2m_Bm_D]}
{[x_1x_2\zeta m_Bm_D+({\bf k}_{1T}-{\bf k}_{2T})^2]
[x_2\zeta m_Bm_D+{\bf k}_{2T}^2]}P_2^\mu\;,
\label{ha}
\end{eqnarray}
with
\begin{eqnarray}
& &\zeta=\eta+\sqrt{\eta^2-1}\nonumber \\
& &\zeta_1=\frac{1}{2}+\frac{\eta-2}{2\sqrt{\eta^2-1}}
\nonumber \\
& &\zeta_2=\eta-1+\frac{2\eta^2-2\eta-1}{2\sqrt{\eta^2-1}}\;.
\end{eqnarray}
The factors $(\not P_1+m_B)\gamma_5/\sqrt{2N_c}$ and
$\gamma_5(\not P_2+m_D)/\sqrt{2N_c}$ come from the matrix
structures of the $B$ and $D$ meson wave functions, respectively.
${\cal C}_F=4/3$ is the color factor, and $N_c$ the number of colors.
Similarly, from fig.~1b we get
\begin{eqnarray}
H^{(b)\mu}&=&
{\rm tr}\left[\gamma_\alpha\frac{\gamma_5(\not P_2+m_D)}{\sqrt{2N_c}}
\gamma^\alpha\frac{\not P_2-\not k_1+m_D}{(P_2-k_1)^2-m_D^2}
\gamma^\mu\frac{(\not P_1+m_B)\gamma_5}{\sqrt{2N_c}}\right]
\nonumber \\
& &\times \frac{-g^2N_c{\cal C}_F}{(k_1-k_2)^2}
\nonumber \\
&=&\frac{16\pi \alpha_s{\cal C}_F[m_D^2+x_1\zeta_2m_Bm_D]}
{[x_1x_2\zeta m_Bm_D+({\bf k}_{1T}-{\bf k}_{2T})^2]
[x_1\zeta m_Bm_D+{\bf k}_{1T}^2]}P_1^\mu
\nonumber \\
& &+\frac{16\pi \alpha_s{\cal C}_F[m_Bm_D-x_1\zeta_1m_B^2]}
{[x_1x_2\zeta m_Bm_D+({\bf k}_{1T}-{\bf k}_{2T})^2]
[x_1\zeta m_Bm_D+{\bf k}_{1T}^2]}P_2^\mu\;.
\label{hb}
\end{eqnarray}
Note that $H^{(b)}$ can be obtained from $H^{(a)}$ by exchanging the
variables associated with the $B$ and $D$ mesons. This permutation symmetry
has been displayed manifestly in fig.~1.

Performing the Fourier transform of
eqs.~(\ref{ha}) and (\ref{hb}) to get ${\tilde H}^\mu$ and substituting
them into (\ref{fbd}), we obtain the factorization formula for
$M^\mu=f_1P_1^\mu+f_2P_2^\mu$, where the form factors $f_1$ and
$f_2$ are given by
\begin{eqnarray}
f_1&=& 16\pi{\cal C}_F\int_{0}^{1}\d x_{1}\d x_{2}\,
\int_{0}^{\infty} b_1\d b_1 b_2\d b_2\,
\phi_B(x_1,b_1)\phi_D(x_2,b_2)
\nonumber \\
& &\times [(m_Bm_D-x_2\zeta_1m_D^2)h(x_1,x_2,b_1,b_2)
\nonumber \\
& &\;\;\;\;+(m_D^2+x_1\zeta_2m_Bm_D)h(x_2,x_1,b_2,b_1)]
\nonumber \\
& &\times \exp[-S(x_i,b_i,m_B,m_D)]\;,
\label{f1}
\end{eqnarray}
and
\begin{eqnarray}
f_2&=& 16\pi{\cal C}_F\int_{0}^{1}\d x_{1}\d x_{2}\,
\int_{0}^{\infty} b_1\d b_1 b_2\d b_2\,
\phi_B(x_1,b_1)\phi_D(x_2,b_2)
\nonumber \\
& &\times [(m_B^2+x_2\zeta_2m_Bm_D)h(x_1,x_2,b_1,b_2)
\nonumber \\
& &\;\;\;\;+(m_Bm_D-x_1\zeta_1m_B^2)h(x_2,x_1,b_2,b_1)]
\nonumber \\
& &\times \exp[-S(x_i,b_i,m_B,m_D)]\;,
\label{f2}
\end{eqnarray}
respectively, with
\begin{eqnarray}
h(x_1,x_2,b_1,b_2)&=&
\alpha_{s}(t)K_{0}\left(\sqrt{x_1x_2\zeta m_Bm_D}b_1\right)
\nonumber \\
& &\times \left[\theta(b_1-b_2)K_0\left(\sqrt{x_2\zeta m_Bm_D}b_1\right)
I_0\left(\sqrt{x_2\zeta m_Bm_D}b_2\right)\right.
\nonumber \\
& &\;\;\;\;\left.
+\theta(b_2-b_1)K_0\left(\sqrt{x_2\zeta m_Bm_D}b_2\right)
I_0\left(\sqrt{x_2\zeta m_Bm_D}b_1\right)\right]\;.
\nonumber\\
& &
\label{dh}
\end{eqnarray}
The wave function $\phi_B$ $(\phi_D)$ comes from ${\cal P}_B$
$({\cal P}_D)$ in eq.~(\ref{fbd}) with the evolution in $P_1^-$ $(P_2^+)$,
which is the result of the resummation of reducible corrections, grouped into
the Sudakov factor. The argument $b$ in $\phi_B$ and $\phi_D$
denotes the intrinsic transverse momentum dependence of the wave
functions \cite{JK}, which is a non-perturbative object, and can
not be handled in perturbation theory.
$K_0$ and $I_0$ are the modified Bessel functions of order zero.
We choose $t$ as the largest scale associated with the hard gluon,
\begin{equation}
t=\max\left(\sqrt{x_1x_2\zeta m_Bm_D},1/b_1,1/b_2\right)\;.
\end{equation}
\vskip 2.0cm

\cl{\large\bf 3. Numerical Results}
\vskip 0.5cm

Before evaluating $f_i$, we compare our formulas with those derived in the
framework of standard PQCD \cite{KK,CM}, where the ${\bf k}_T$ dependence
in the hard scattering amplitude is neglected, and the heavy meson
wave functions, with ${\bf k}_T$ integrated, take the simple form of the
$\delta$-function (the so-called peaking approximation),
\begin{equation}
\phi_B(x)=\frac{f_B}{2\sqrt{3}}\delta(x-x_B),\;\;\;\;\;
\phi_D(x)=\frac{f_D}{2\sqrt{3}}\delta(x-x_D)\;.
\label{pa}
\end{equation}
Here $f_B=0.12$ GeV and $f_D=0.14$ GeV are the decay
constants of the $B$ and $D$ mesons \cite{BLS}, respectively.
Eqs.~(\ref{f1}) and (\ref{f2}) are then reduced to the standard factorization
formulas without $b$ integrations, which lead to
\begin{eqnarray}
f_1&=& \frac{4}{3}\pi{\cal C}_F\alpha_sf_Bf_D
\left[\frac{m_Bm_D-x_D\zeta_1m_D^2}{x_Bx_D^2\zeta^2m_B^2m_D^2}+
\frac{m_D^2+x_B\zeta_2m_Bm_D}{x_B^2x_D\zeta^2m_B^2m_D^2}\right]\;,
\nonumber \\
f_2&=& \frac{4}{3}\pi{\cal C}_F\alpha_sf_Bf_D
\left[\frac{m_B^2+x_D\zeta_2m_Bm_D}{x_Bx_D^2\zeta^2m_B^2m_D^2}+
\frac{m_Bm_D-x_B\zeta_1m_B^2}{x_B^2x_D\zeta^2m_B^2m_D^2}\right]\;.
\label{f222}
\end{eqnarray}

It is apparent that the above expressions
are very sensitive to the values of $x_B$ and $x_D$, and the coupling
constant $\alpha_s$ must be regarded as a free parameter. We consider the
non-leptonic decay $B\to D\pi$, which corresponds to
the case of maximum recoil here with $\eta=
\eta_{\rm max}=1.59$. Setting $\alpha_s=0.4$, $x_B=0.07$ and $x_D=0.2$
as in \cite{CM}, we obtain $f_1+f_2=1.3$, which gives a branching
ratio comparable with experimental data \cite{KBS}. However, if
slightly different values like $x_B=0.07$ and $x_D=0.15$
were inserted, the branching ratio becomes 3 times larger. On the other
hand, simply setting $\alpha_s$ to a constant makes the justification
of the perturbative calculation unavailable. Compared to the standard
PQCD approach, our modified perturbative expressions
are less sensitive to the profile change of the wave functions
due to the inclusion of ${\bf k}_T$ in the hard scattering amplitude,
which moderates the divergences from small $x_B$ and $x_D$.
Substituting eq.~(\ref{pa}) into (\ref{f1}) and (\ref{f2}), and performimg
the integrations over $b_1$ and $b_2$, we find that
the latter set of $x_B$ and $x_D$ leads to a branching ratio
only 50\% larger than that from the former set.

We adopt the following model for the $B$ meson wave function \cite{S},
\begin{equation}
\Phi_B(x,{\bf k}_T)=N_B\left[C_B+\frac{m_B^2}{1-x}+\frac{{\bf k}_T^2}
{x(1-x)}\right]^{-2}\;.
\label{bwf}
\end{equation}
The constants $N_B$ and $C_B$ are determined by the normalizations
\begin{eqnarray}
& &\int_0^1\d x\int\d^2 {\bf k}_T\Phi_B(x,{\bf k}_T)=\frac{f_B}{2\sqrt{3}} \;,
\nonumber \\
& &\int_0^1\d x\int\d^2 {\bf k}_T[\Phi_B(x,{\bf k}_T)]^2=\frac{1}{2}\;,
\end{eqnarray}
which give $N_B=0.923$ GeV$^3$ and $C_B=-27.877255$ GeV$^2$.
$\phi_B$ is then defined by
\begin{eqnarray}
\phi_B(x,b)&=&\int\d^2 {\bf k}_T\Phi_B(x,{\bf k}_T)
e^{i{\bf k}_T\cdot {\bf b}}
\nonumber \\
&=&\frac{\pi N_B bx^2(1-x)^2}{\sqrt{m_B^2x+C_Bx(1-x)}}
K_1\left(\sqrt{m_B^2x+C_Bx(1-x)}b\right)\;.
\label{bwfb}
\end{eqnarray}
It is observed that $\phi_B$ peaks at $x\approx 0$, and decreases monotonically
with $x$ for a fixed $b$, signifying the soft dynamics involved in the
rest $B$ meson.

If assuming the similar model for the $D$ meson wave function
with $m_B$ in eq.~(\ref{bwf}) replaced by $m_D$ straightforwardly,
\begin{equation}
\phi_D(x,b)=\frac{\pi N_D bx^2(1-x)^2}{\sqrt{m_D^2x+C_Dx(1-x)}}
K_1\left(\sqrt{m_D^2x+C_Dx(1-x)}b\right)\;,
\label{dwfb}
\end{equation}
we obtain the constants
$N_D=0.136$ GeV$^3$ and $C_D=-3.495345$ GeV$^2$. The resulting wave function
$\phi_D$ also peaks at small $x\approx 0.01$ for a fixed $b$.
However, the QCD sum rule analysis in \cite{CZ} has shown that
the average momentum fraction of the light valence quark in a fast $D$ meson
is roughly 0.2.
To be consistent with this observation, we employ eq.~(\ref{dwfb})
but with $C_D$ determined by the requirement that $\phi_D$ takes the
maximum value at $x\approx 0.2$ for $b\to 0$.
We then have $C_D=-2.9$ GeV$^2$, along with
$N_D=0.240$ GeV$^3$ from the normalization
$\int \d x\phi_D(x,0)=f_D/(2\sqrt{3})$.

Results of $f_1$ and $f_2$ derived from eqs.~(\ref{f1}) and (\ref{f2}),
respectively, with $b_1$ and $b_2$ integrated up to the same
cutoff $b_c$ are shown in fig.~2.
We find that at $\eta=1.30$ approximately 55\% of the contribution
to $f_i$ comes from the region with $\alpha_s(1/b_c)< 1$, or equivalently,
$b_c< 0.5/\Lambda$. The percentage of perturbative contribution increases
with $\eta$, and for $\eta$ above 1.39, more than 60\% of the full
contribution is accumulated in this region.
It implies that our PQCD analysis of the decay $B\to D l \nu$
in the range of $\eta \ge 1.39$
is relatively reliable, since perturbative contribution dominates \cite{LS}.
It is also found that the self-consistency of the perturbation theory
becomes worse quickly for $\eta < 1.3$ as expected.

Based on the above conclusion, we are led to consider the
two-body non-leptonic decays
such as $B\to D\pi$ and $B\to DD_s$, which can be best described by our PQCD
formalism. The decay rate of the specific mode ${\bar B}^0\to D^+\pi^-$
is given by
\begin{equation}
\Gamma=\frac{1}{64\pi}G_F^2|V_{ud}|^2|V_{cb}|^2f_\pi^2m_B^3
\left(1-\frac{m_D^2}{m_B^2}\right)^3|f_1+f_2|^2\;,
\label{dp}
\end{equation}
which is derived from the amplitude
\begin{equation}
A=\frac{G_F}{\sqrt{2}}V_{ud}V_{cb}\langle\pi|\gamma_{\mu}(1-\gamma_5)
|0\rangle\langle D|{\bar c}\gamma^{\mu}b|B\rangle
\label{a}
\end{equation}
with the PCAC relation $\langle\pi(P)|\gamma_{\mu}(1-\gamma_5)
|0\rangle=i\sqrt{2}f_\pi P_\mu$ inserted, $f_\pi=93$ MeV being the pion
decay constant. Eq.~(\ref{a}) is achieved following the conclusion in
\cite{KP} that the non-factorizable $W$-exchange contribution is negligible.
The value of $f_1+f_2$ in this case can be easily
read off from the curves corresponding to $\eta=1.59$ in
fig.~2, which is equal to 1.44. Substituting the matrix element
$|V_{ud}|=0.974$, we obtain $\Gamma=8.4\times 10^{-13}|V_{cb}|^2$ GeV,
or equivalently, the branching ratio $B({\bar B}^0\to D^+\pi^-)=
1.65|V_{cb}|^2$ from the total width $(0.51\pm 0.02)\times
10^{-9}$ MeV of the ${\bar B}^0$ meson \cite{RPP}.
Comparing with experimental data $B({\bar B}^0\to D^+\pi^-)=
3.2\times 10^{-3}$, we extract the matrix element
$|V_{cb}|=0.044$, consistent with currently accepted value \cite{RPP}.
Similarly, the decay rate for the mode ${\bar B}^0\to D^+D_s^-$ is given by
\begin{eqnarray}
\Gamma&=&\frac{1}{32\pi}G_F^2|V_{cs}|^2|V_{cb}|^2f_{D_s}^2\frac{m_D}{m_B^2}
\sqrt{\eta^{'2}_{\rm max}-1}
\nonumber \\
& &\times\left|(m_B^2-m_D^2+m_{D_s}^2)f_1+
(m_B^2-m_D^2-m_{D_s}^2)f_2\right|^2
\label{ddp}
\end{eqnarray}
with the matrix element $|V_{cs}|=1.0$, the decay constant of the
$D_s$ meson $f_{D_s}=0.16$ GeV \cite{BLS}, and the $D_s$
meson mass $m_{D_s}=1.97$ GeV.
In this case we have the maximum velocity transfer
$\eta'_{\rm max}=(m_B^2+m_D^2-m_{D_s}^2)/(2m_Bm_D)=1.39$, for which the
corresponding values $f_1=0.47$ and $f_2=1.32$ are read off from fig.~2.
Eq.~(\ref{ddp}) then gives the decay rate $\Gamma=2.7\times 10^{-12}
|V_{cb}|^2$, or the branching ratio
$B({\bar B}^0\to D^+D_s^-)=5.3|V_{cb}|^2$. Experimental data show
$B({\bar B}^0\to D^+D_s^-)=9.9\times 10^{-3}$,
from which we extract $|V_{cb}|=0.043$, close to that obtained from
the decay ${\bar B}^0\to D^+\pi^-$.

Due to the consistency of our predictions with experimental data,
we can explore the behavior
of the IW function near the high end of $\eta$ reliably.
For finite $m_B$ and $m_D$, eq.~(\ref{iw}) is modified to
\begin{equation}
M^\mu=\sqrt{m_Bm_D}(\xi_+(v_1\cdot v_2)(v_1+v_2)^\mu+
\xi_-(v_1\cdot v_2)(v_1-v_2)^\mu)\;,
\label{iwm}
\end{equation}
where $\xi_+\to \xi$ and $\xi_-\to 0$ in the heavy
meson limit. A simple manipulation gives the relations
\begin{equation}
\xi_{\pm}=\frac{1}{2}\left(\sqrt{\frac{m_B}{m_D}}f_1 \pm
\sqrt{\frac{m_D}{m_B}}f_2\right)\;.
\end{equation}
The dependence of $\xi_+$ and $\xi_-$ on $\eta$ is shown in fig.~3,
which exhibits a falloff and an increase with $\eta$, respectively.
The magnitude of $\xi_-$ indeed diminishes as stated above.
A model calculation of the IW function in terms of the overlap integrals
of the heavy meson wave functions has been performed \cite{N3}, which leads to
\begin{equation}
\xi_m(\eta)=\frac{2}{\eta+1}\exp\left[-(2\rho^2-1)\frac{\eta-1}{\eta+1}
\right]
\label{ov}
\end{equation}
with the parameter $\rho\approx 1$. The behavior of $\xi_m$
is also shown in fig.~3. It is observed that our predictions for $\xi_+$
are close to $\xi_m$ at large $\eta$, and begin to deviate from $\xi_m$
as $\eta < 1.39$. The match confirms the applicability of PQCD to
heavy meson decays in the large recoil region.

At last, the differential decay rate for the specific mode
${\bar B}^0\to D^+ l^-{\bar \nu}$ with vanishing
lepton masses is given by
\begin{eqnarray}
\frac{\d \Gamma}{\d \eta}\equiv |V_{cb}|^2R(\eta)
=|V_{cb}|^2\frac{1}{48\pi^3}G_F^2m_Bm_D^4
(\eta^2-1)^{3/2}|f_1+f_2|^2\;.
\label{dd}
\end{eqnarray}
Substituting the results of $f_i$ into eq.~(\ref{dd}), we derive
the behavior of $R(\eta)$ for $\eta \ge 1.39$ as in fig.~4,
which shows an increase with $\eta$. Once experimental data for the
spectrum of the decay ${\bar B}^0\to D^+ l^-{\bar \nu}$ are available, the
matrix element $|V_{cb}|$ can also be extracted from the curve in fig.~4.

\vskip 2.0cm

\cl{\large \bf 4. Conclusions}
\vskip 0.5cm

In this paper we have applied the PQCD formalism to the
semi-leptonic decay $B\to D l\nu$, and found that the perturbative
calculation is reliable for the velocity transfer above 1.4.
The point is to include the resummation of large radiative corrections
in the process, which improves the
applicability of PQCD. The intrinsic
transverse momentum dependence also plays an essential role in the
calculation. We emphasize that our analysis
does not involve any phenomenological parameter, and is insensitive to
the profile change of the wave functions. The perturbative calculation
is justified by considering the magnitude of
the running coupling constant, which defines
the region where the perturbation theory is reliable.

Our predictions are satisfying in the sense that they match
the model estimation of the IW function at the high end of velocity transfer,
and the values 0.044 and 0.043 for the matrix
element $|V_{cb}|$ are extracted
from the decays $B\to D\pi$ and $B\to DD_s$, respectively. The results
presented in this work confirm our perturbative analysis of the decay
$B\to \pi l\nu$ in \cite{LY}, which is important for the extraction of
the matrix element $|V_{ub}|$.

\vskip 0.5cm
I thank F.T. Cheng, T. Huang, G. Sterman and H.L. Yu for helpful discussions.
This work was supported by the National
Science Council of R.O.C. under Grant No. NSC84-2112-M194-006.
\vskip 2.0cm

\centerline{\large \bf Appendix}
\vskip 0.3cm

In this appendix we show the derivation of the exponent $s(x,b,Q)$
in eq.~(\ref{sdb}).
We start with eq.~(5.42) in ref.~\cite{BS}
\begin{equation}
s(x,b,Q)=\int_{1/b}^{xQ}\frac{\d \mu}{\mu}\left[\ln\left(\frac{xQ}{\mu}
\right)A(g(\mu))+B(g(\mu))\right]\;,
\label{fsl}
\end{equation}
in which the factors $A(g)$ and $B(g)$ are expanded as
\begin{eqnarray}
A(g)&=&A^{(1)}\frac{\alpha_s}{\pi}+A^{(2)}
\left(\frac{\alpha_s}{\pi}\right)^2
\nonumber \\
B(g)&=&\frac{2}{3}\frac{\alpha_s}{\pi}\ln\left(\frac{e^{2\gamma-1}}
{2}\right)\;,
\end{eqnarray}
in order to take into account the next-to-leading logarithms.
The running coupling constant $\alpha_s$ is written as
\begin{equation}
\frac{\alpha_s(\mu)}{\pi}=\frac{1}{\beta_1\ln(\mu^2/\Lambda^2)}-
\frac{\beta_2}{\beta_1^3}\frac{\ln\ln(\mu^2/\Lambda^2)}
{\ln^2(\mu^2/\Lambda^2)}\;.
\end{equation}
The above coefficients $\beta_{i}$ and $A^{(i)}$ are
\begin{eqnarray}
& &\beta_{1}=\frac{33-2n_{f}}{12}\;,\;\;\;\beta_{2}=\frac{153-19n_{f}}{24}\; ,
\nonumber \\
& &A^{(1)}=\frac{4}{3}\;,
\;\;\; A^{(2)}=\frac{67}{9}-\frac{\pi^{2}}{3}-\frac{10}{27}n_
{f}+\frac{8}{3}\beta_{1}\ln\left(\frac{e^{\gamma}}{2}\right)\; ,
\label{12}
\end{eqnarray}
where $n_f=4$ is the number of quark flavors, and
$\gamma$ is the Euler constant. Performing the integration in
eq.~(\ref{fsl}), we obtain $s$, which
is given in terms of the variables
\begin{eqnarray}
{\hat q} \equiv  {\rm ln}\left(x Q/\Lambda\right),\;\;\;\;\;
{\hat b} \equiv {\rm ln}(1/b\Lambda)
\label{11}
\end{eqnarray}
by \cite{LS}
\begin{eqnarray}
s&=&\frac{A^{(1)}}{2\beta_{1}}\hat{q}\ln\left(\frac{\hat{q}}
{\hat{b}}\right)-
\frac{A^{(1)}}{2\beta_{1}}\left(\hat{q}-\hat{b}\right)+
\frac{A^{(2)}}{4\beta_{1}^{2}}\left(\frac{\hat{q}}{\hat{b}}-1\right)
\nonumber \\
& &-\left[\frac{A^{(2)}}{4\beta_{1}^{2}}-\frac{A^{(1)}}{4\beta_{1}}
\ln\left(\frac{e^{2\gamma-1}}{2}\right)\right]
\ln\left(\frac{\hat{q}}{\hat{b}}\right)
\nonumber \\
& &+\frac{A^{(1)}\beta_{2}}{4\beta_{1}^{3}}\hat{q}\left[
\frac{\ln(2\hat{q})+1}{\hat{q}}-\frac{\ln(2\hat{b})+1}{\hat{b}}\right]
\nonumber \\
& &+\frac{A^{(1)}\beta_{2}}{8\beta_{1}^{3}}\left[
\ln^{2}(2\hat{q})-\ln^{2}(2\hat{b})\right]
\nonumber \\
& &+\frac{A^{(1)}\beta_{2}}{8\beta_{1}^{3}}
\ln\left(\frac{e^{2\gamma-1}}{2}\right)\left[
\frac{\ln(2\hat{q})+1}{\hat{q}}-\frac{\ln(2\hat{b})+1}{\hat{b}}\right]
\nonumber \\
& &-\frac{A^{(1)}\beta_{2}}{16\beta_{1}^{4}}\left[
\frac{2\ln(2\hat{q})+3}{\hat{q}}-\frac{2\ln(2\hat{b})+3}{\hat{b}}\right]
\nonumber \\
& &-\frac{A^{(1)}\beta_{2}}{16\beta_{1}^{4}}
\frac{\hat{q}-\hat{b}}{\hat{b}^2}\left[2\ln(2\hat{b})+1\right]
\nonumber \\
& &+\frac{A^{(2)}\beta_{2}^2}{1728\beta_{1}^{6}}\left[
\frac{18\ln^2(2\hat{q})+30\ln(2\hat{q})+19}{\hat{q}^2}
-\frac{18\ln^2(2\hat{b})+30\ln(2\hat{b})+19}{\hat{b}^2}\right]
\nonumber \\
& &+\frac{A^{(2)}\beta_{2}^2}{432\beta_{1}^{6}}
\frac{\hat{q}-\hat{b}}{\hat{b}^3}
\left[9\ln^2(2\hat{b})+6\ln(2\hat{b})+2\right]\;.
\label{sss}
\end{eqnarray}
The previous studies involving the Sudakov logarithms pick up only
the first six terms in eq.~(\ref{sss}), which are more important
than the remaining ones in large $Q$ region.
Note that the coefficients of
the fifth and sixth terms are different from those in refs.~\cite{BS,LS}. It
can be easily checked that with these corrections the results
for the pion form factor in \cite{LS} are reducend only by few percents.
An explicit examination on the form factors $f_i$ in $B\to D$ decays
shows that the partial expression
including only the first six terms give predictions
smaller than those from the full
expression by less than 5\%. Hence, for simplicity this
partial expression is substituted into (\ref{f1})
and (\ref{f2}). Note that $s$ is defined for
${\hat q}\ge {\hat b}$, and is set to zero for ${\hat q}<{\hat b}$.
As a similar treatment, the complete Sudakov factor $\exp(-S)$ is set to
unity, if $\exp(-S)>1$, during the numerical analysis.

\newpage
\cl{\large \bf Figure Captions}
\vskip 0.5cm

\noindent
{\bf Fig. 1.} Lowest-order diagrams of the decay $B\to D l\nu$.
\vskip 0.5cm

\noindent
{\bf Fig. 2.} Dependence of (a) $f_1$ and of (b) $f_2$
on the cutoff $b_c$  for (1) $\eta=1.3$,
(2) $\eta=1.39$, and (3) $\eta=1.59$.
\vskip 0.5cm

\noindent
{\bf Fig. 3.} Dependence of (a) $\xi_+$ and of (b) $\xi_-$
on $\eta$ derived from our PQCD formalism.
The dependence of $\xi_m$ on $\eta$
from the model calculation in \cite{N3} (dashed line) is also shown.
\vskip 0.5cm

\noindent
{\bf Fig. 4.} Dependence of $R(\eta)$ on $\eta$ derived from
the PQCD analysis.

\end{document}